# Gap reopening as a possible signature of coupling between Majorana zero modes in $Sn-(Bi,Sb)_2(Te,S)_3$-based Josephson trijunctions

Duolin Wang[1,2], Xiang Zhang[1], Yunxiao Zhang[1,2], Heng Zhang[3], Fucong Fei[3], Xiang Wang[1,2], Bing Li[1,2], Xiaozhou Yang[1,2], Yukun Shi[1,2], Zhongmou Jia[1,2], Enna Zhuo[1,2], Yuyang Huang[1,2], Anqi Wang[1,2], Zenan Shi[1,2], Zhaozheng Lyu[1,2,4,†], Xiaohui Song[1,4], Peiling Li[1], Bingbing Tong[1], Ziwei Dou[1], Jie Shen[1], Guangtong Liu[1,4], Fanming Qu[1,2,4] Fengqi Song[3,†] and Li Lu[1,2,4,†]

[1] *Beijing National Laboratory for Condensed Matter Physics, Institute of Physics, Chinese Academy of Sciences, Beijing 100190, China*

[2] *School of Physical Sciences, University of Chinese Academy of Sciences, Beijing 100049, China*

[3] *College of Physics, Nanjing University, Nanjing 210008, China*

[4] *Hefei National Laboratory, Hefei 230088, China*

† Corresponding authors: lyuzhzh@iphy.ac.cn, songfengqi@nju.edu.cn, lilu@iphy.ac.cn

**Abstract**

In the past two decades, enormous efforts have been made to search for possible platforms and schemes to implement topological quantum computation (TQC). In exploring the Fu-Kane scheme of TQC based on Josephson trijunctions constructed on topological insulators, the predicted Majorana phase diagram of an individual trijunction has already been verified experimentally. If Majorana zero modes indeed exist in this kind of trijunctions, coupling between them in multiple trijunction devices should be further expected. In this study, we fabricated Josephson devices containing two adjacent Josephson trijunctions on the surface of $Sn-(Bi,Sb)_2(Te,S)_3$, and observed a possible signature of the coupling effect manifested as the reopening of minigap in both trijunctions where a closure would otherwise be expected if the trijunctions existed individually. While alternative interpretations cannot be fully ruled out, our findings provide new experimental support for the validity of the Fu-Kane theory and provide further motivation for advancing the TQC scheme proposed by Fu and Kane.

Majorana zero mode (MZM) is the key ingredient for implementing fault-tolerant topological quantum computation (TQC) [1-3]. Direct or indirect evidence for the existence of MZMs in condensed matter systems have been obtained, including the zero-bias conductance peaks (ZBCPs) in nanowire devices [4-6], atomic chains [7], artificial chains [8,9], vortex cores of topological insulator (TI)-based [10,11] and iron-based superconductors [12-14], as well as the $4\pi$-period current-phase relation in Josephson junctions based on nanowires [15] and TIs [16,17]. Recently, efforts have further been made along the direction of hosting multiple MZMs and coupling them with high-frequency circuits, aiming for further braiding/fusing the MZMs and for single-shot readout of the parity of the systems [18,19].

Compared to the nanowire-based scheme of TQC [20] which relies on an array of nanowires, the Fu-Kane scheme of TQC has the advantages of using three-dimensional TI-based Josephson trijunctions as the building blocks to form a scalable topological quantum circuits on the TI surfaces [21], and further employing surface code technique to realize universal TQC [22]. In this scheme, according to Fu and Kane [21], there will be a local Majorana bound state associated with a MZM nucleated at the center of the trijunction when the minigaps in odd numbers of single junctions are negative. Here, the minigap is defined as half of the level spacing between electron-like and hole-like Andreev bound states (ABSs). Previously, Fu-Kane's MZM phase diagram for single Josephson trijunctions constructed on $Bi_2Te_3$ surface has been experimentally verified [23].

In order to further explore the Fu-Kane scheme of TQC, in this work we study whether the MZMs in multiple trijunctions can couple with each other, just as it happened in nanowire devices [24] and vortex cores [25]. We use the term 'coupled' rather than 'fused,' as the current experimental setup does not permit parity readout to determine whether the interaction of two MZMs results in a fermion or the vacuum state [1-3]. We solely focus on the expected reopening of closed minigaps induced by MZM coupling.

Figure 1 shows the devices we designed and fabricated, which contain two adjacent Josephson trijunctions on the surface of a $Sn-(Bi,Sb)_2(Te,S)_3$ flake. Four zigzag-shaped superconducting electrodes (50-nm-thick Al with a fully oxidized surface) form five single Josephson junctions, each spaced approximately 180 nm apart (defining the length of the junctions). The single junctions located at the upper left, lower left, upper right, and lower right have a width of 2 μm, whereas the internal single junction connecting the two trijunctions has a width (i.e., an inter-trijunction distance) of either 1 μm or 2 μm. The six out of eight terminals of the four superconducting electrodes are interconnected to form three flux loops with areas of $S_1$, $S_2$ and $S_3$, enabling adjustment of the phase difference in each embedded single junction through application of a global magnetic field. And, to detect the local minigaps and thereby obtain the coupling effect between the two trijunctions, two normal-metal electrodes (2 nm Al followed by 120 nm Au) are introduced to contact only at the centers of the trijunctions. Previously, it has been shown that the contact conductance contains the information of local minigap [23,26,27].

For the devices shown in Fig. 1, the areas of the three flux loops satisfy $S_1 = 2S_2 = 2S_3$. With this areal configuration, and if there is no coupling between the two trijunctions, the status of both trijunction should evolve independently along the diagonal line in Fig. 2(a) when a global magnetic field is applied and varied, just what happened in previously studies [23, 26]. More specifically, the status of the upper trijunction probed by the blue electrode should evolve in the phase space with blue coordinates spanned by $\varphi_1$ (the phase difference of the upper left single junction determined by the flux in $S_1$) and $\varphi_2$+$\varphi_3$ (the phase difference of the internal single junction determined by the flux in $S_2 + S_3$). The minigap at the center should open in the white regions and close in the light blue regions. Simultaneously, the status of the lower trijunction probed by the red electrode should evolve in the space with red coordinates spanned by $\varphi_2$ (the phase difference of the lower left single junction determined by the flux in $S_2$) and $\varphi_3$ (the phase difference of the lower right single junction determined by the flux in $S_3$). The minigap should open in the white regions and close in the light red regions. Figures 2(b, c) depict the qualitative line shapes of minigap in the upper and the lower trijunction's centers in the absence of coupling between them. Within the parametric ranges plotted, the minigap of the upper trijunction should oscillate between opening and closing twice, and the minigap of the lower trijunction should oscillate between opening and closing once.

In Fu-Kane's model, the trijunction is treated as three single junctions with infinite width. In real experiments, however, the width of single junctions is always finite (ranging from 1 to 2 μm in this experiment). It is already known that the topological boundary states in this type of devices are not so local down to a micron length scale [27, 28]. Therefore, the measured contact conductance at one trijunction's center will inevitably contain to a certain extent the information of the density of states of the other trijunction's center. Figures 2(d, e) depicts the modified qualitative line shapes of minigap in the absence of MZM coupling, but after considered state penetration from one trijunction's center to another, assuming that the penetration causes 1/4 change in the apparent amplitude of the measured minigap.

According to the Fu-Kane model for single trijunction, both trijunctions undergo minigap closing and MZM occurring in the purple-colored regions in Fig. 2(a). If there is no coupling between the two MZMs, the minigaps at both centers should keep closed even with state penetration from one to the other. If there is coupling, however, the minigap at both centers will reopen, which will turn the local minima in Figs. 2(d, e) into local maxima, as depicted in Figs. 2(f, g). These expected line shapes were indeed observed in this experiment.

Figures 2(h, l) are the 2D color maps of contact resistance measured by the blue and the red probe electrodes, respectively, as a function of bias current $I_{\mathrm{bias}}$ and global magnetic field $B$. The apparent oscillation period in Fig. 2(h) is $2T_1 = 0.47$ G, with $T_1$ in agreement with the period expected from the loop area $S_1$: $T_1 = \phi_0/S_1 = 0.23$ G (where $\phi_0$ is the flux quantum and $S_1 \approx 90\ \mu\mathrm{m}^2$). Strikingly, the pattern demonstrates three lobes within the interval of $2T_1$. We will discuss the origin of this phenomenon later. For the data measured by the red electrode and displayed in Fig. 2(l), the apparent oscillation period $T_2 = 0.45$ G is in agreement with the expected period $\phi_0/S_2 = 0.44$ G. However, there are internal structures within each period, which will also be discussed later.

Let us discuss in details the data measured by the blue electrode. The upper curve in Fig. 2(i) is the magnetic field dependence of $(\mathrm{d}V/\mathrm{d}I_{\mathrm{bias}})^{-1}$, the inverse of contact resistance in Fig. 2(h) along the horizontal linecut at $I_{\mathrm{bias}} = -70$ nA. Although the amplitude of $(\mathrm{d}V/\mathrm{d}I_{\mathrm{bias}})^{-1}$ oscillation is only ~7%, it actually reflects a considerable oscillation of the minigap. To support this argument, we convert the vertical linecuts, namely the $\mathrm{d}V/\mathrm{d}I_{\mathrm{bias}} - I_{\mathrm{bias}}$ curves, at each magnetic field in Fig. 2(h) to $V - I_{\mathrm{bias}}$ curves through integration, and then further convert them to $\mathrm{d}I/\mathrm{d}V_{\mathrm{bias}} - V_{\mathrm{bias}}$ curves through differential. One of the converted curves is shown in Fig. 2(j). It can be seen that there are two big peaks representing the induced gap at approximately $\pm 130$ μeV, plus two small peaks representing the ABS-related minigap whose position oscillates significantly with the variation of magnetic field [the lower curve in Fig. 2(i)]. It demonstrates three lobes in two periods [Fig. 2(k)], which is in agreement with the expected qualitative line shape depicted in Fig. 2(f), supporting that the minigap reopens in the purple regions of the phase diagram.

For the data taken by the red electrode at the lower trijunction's center and displayed in Fig. 2(l), the quality of the data is not as high as those in Fig. 2(h), such that the minigap cannot be directly read out from the $\mathrm{d}I/\mathrm{d}V_{\mathrm{bias}} - V_{\mathrm{bias}}$ curves. Nevertheless, previous studies have established that there exists a correspondence between the contact conductance and the extracted minigap in their magnetic field dependences [23,26,27,29]. This correspondence is also true for the data measured by the blue electrode in this experiment, as shown in Fig. 2(i). Following this correspondence, in Fig. 2(m) we use the normalized contact resistance along the horizontal linecut at $I_{\mathrm{bias}} = -15$ nA in Fig. 2(l) to represent the magnetic field dependence of the minigap at the lower trijunction's center. To improve the signal-to-noise ratio and obtain the red line in Fig. 2(m), the data points within central six periods [not all shown in Fig. 2 (l)] are folded into one period and then averaged. The overall line shape of the data is again in agreement with the expected qualitative line shape of minigap depicted in Fig. 2(g), supporting that the minigap at the lower trijunction's center reopens and maximizes in the purple-colored regions of the phase diagram.

For comparison, we have also studied similar devices as shown in Fig. 1, but with equal loop areas $S_1 = S_2 = S_3 = 45$ μm$^2$. The phase diagrams of these control devices are shown in Fig. 3(a). When a global magnetic field is applied, the status of the upper trijunction's center will evolve along the blue line in the phase space spanned by the blue coordinates, and the status of the bottom trijunction will evolve simultaneously along the red line in the phase space spanned by the red coordinates. Figures 3(b, c) depict the qualitative line shapes of magnetic field dependence of minigap at the upper and lower trijunction's centers, respectively. The minigap at the upper trijunction's center is expected to oscillate twice within an interval of 2π, whereas the minigap at the lower trijunction's center is expected to oscillate only once within the same interval. Figures 3(d, e) further depict the qualitative line shapes of minigap after considered state penetration from one center to the other, for the reasons explained before. Overall, in this type of devices there will be no coexistence of MZM, thus no reopening of minigap caused by MZM coupling.

About ten control devices were fabricated and investigated. Figures 3(f, g) present the typical 2D color maps of contact resistance measured by the blue and the red probe electrodes on one of them, respectively, as a function of $I_{\mathrm{bias}}$ and $B$. Figures 3(h, i) further present the magnetic field dependences of $(\mathrm{d}V/\mathrm{d}I_{\mathrm{bias}})^{-1}$, the inverse of the data along horizontal linecuts at 10 nA in Fig. 3(f) and at 30 nA in Fig. 3(g), respectively. According to the contact conductance-minigap correspondence, the oscillations of $(\mathrm{d}V/\mathrm{d}I_{\mathrm{bias}})^{-1}$ in Figs. 3(h, i) represent the oscillation of local minigap. Also superimposed in Figs. 3(h, i) are two periods of expected qualitative line shapes shown in Figs. 3(d, e), with the gray regions denoting the minigap closure state. Obviously, there is qualitative agreement between the data and the expected line shapes. The results indicate that, in the $S_1 = S_2 = S_3$ configuration where MZM does not coexist during the variation of a global magnetic field, the minigap in each trijunction evolves independently as in an individual trijunction.

In the above, we have assumed and explained that the states (i.e., the wavefunctions) at one trijunction's center can penetrate to the other trijunction's center. To further verify the existence of state penetration, we studied a different type of device as shown in Fig. 4(a). Three superconducting wires made of Al were added to allow independent local control of magnetic flux in the three superconducting loops. We first verified on two individual devices that the contact resistance (thus the minigap) at the lower trijunctions' centers was sensitive to the variation of global magnetic field [Figs. 4(b, d)]. Then, we switched to local flux control [Figs. 4(c, e)]. A rough estimation based on the Biot-Savart law shows that the magnetic flux brought by $I_1$ in the lower left or lower right loops will be roughly an order of magnitude smaller than that brought by $I_2$ or $I_3$, if $I_1 = I_2 = I_3$. It is therefore expected that current $I_1$ should mainly adjust the phases of the two upper single junctions, it should not affect the lower trijunction much via flux crosstalk. And similar expectation applies to currents $I_2$ and $I_3$. However, Figs. 4(c, e) show that $I_1$ influenced the lower trijunction's state as significantly as $I_2$ and $I_3$ did, and vice versa - a phenomenon cannot be fully attributed to flux crosstalk. The results reflect that the upper trijunction's wavefunction penetrates to the lower one's center to a certain extent, and vice versa. This conclusion is in agreement with our previous studies on Josephson devices based on similar TIs [23,27,28].

Finally, let us address some of the limitations and open questions of this experiment. First, we observed the reopening of minigap on only two out of more than ten devices. An example of without reopening is presented in the Supplemental Material [30]. The absence of coupling effect in most devices might reflect the overall low coherence length of the MZMs, which would be caused by the not-high-enough quality of the TI material. We notice that the coupling of MZMs in nanowire devices often suffers from a similar problem. Second, because of the low success rate in observing the coupling effect, we were unable to verify whether the size of the reopened minigap after MZMs coupling meets the theoretical distance-dependent formula given in Ref. [31]. So far, only qualitative evidences were obtained in this experiment. Third, the assumed state penetration from one trijunction's center to the other and its phenomenological influence on the amplitude of the minigap needs more rigorous theoretical studies. The single Josephson junctions in the devices are in the small-junction limit [32] compared to the Josephson penetration depth, forming a nonlocal network. The analysis of such a complex network is beyond our theoretical capabilities. Lastly, how to obtain the fusion result of the MZMs through rapid measurement and further verify their non-Abelian statistical properties remains a rather distant goal.

In conclusion, we have investigated the coupling effect between MZMs in TI-based Josephson devices containing two trijunctions, and found that the magnetic field dependent line shapes of the minigaps at the centers of the trijunctions are in qualitative agreement with the anticipated ones when the MZMs in the two trijunctions are coupled with each other. Given the experimental limitations discussed above, we regard our observations as a possible signature of coupling between MZMs rather than a definitive demonstration. Nevertheless, our work would further validate the correctness of the Fu-Kane theory from the experimental side, motivating continued advancement along the TQC scheme proposed by Fu and Kane.

**Acknowledgements**

This work was supported by the Innovation Program for Quantum Science and Technology through Grant No. 2021ZD0302600; by NSFC through Grant Nos. 92065203, 92365302, 11527806, 12074417, 11874406, 11774405, E2J1141, 92161201, 12374043 and 12474272; by the Strategic Priority Research Program B of the Chinese Academy of Sciences through Grants Nos. XDB33010300, XDB28000000, and XDB07010100; by the National Basic Research Program of China through MOST Grant Nos. 2016YFA0300601, 2017YFA0304700, 2015CB921402 and 2022YFA1402404; by Beijing Natural Science Foundation through Grant No. JQ23022; by Beijing Nova Program through Grant No. Z211100002121144; and by Synergetic Extreme Condition User Facility (SECUF).

**Data Availability**

The data that support the findings of this study are openly available in Zenodo at 10.5281/zenodo.20427617, see Ref. [33].

## Figure Captions

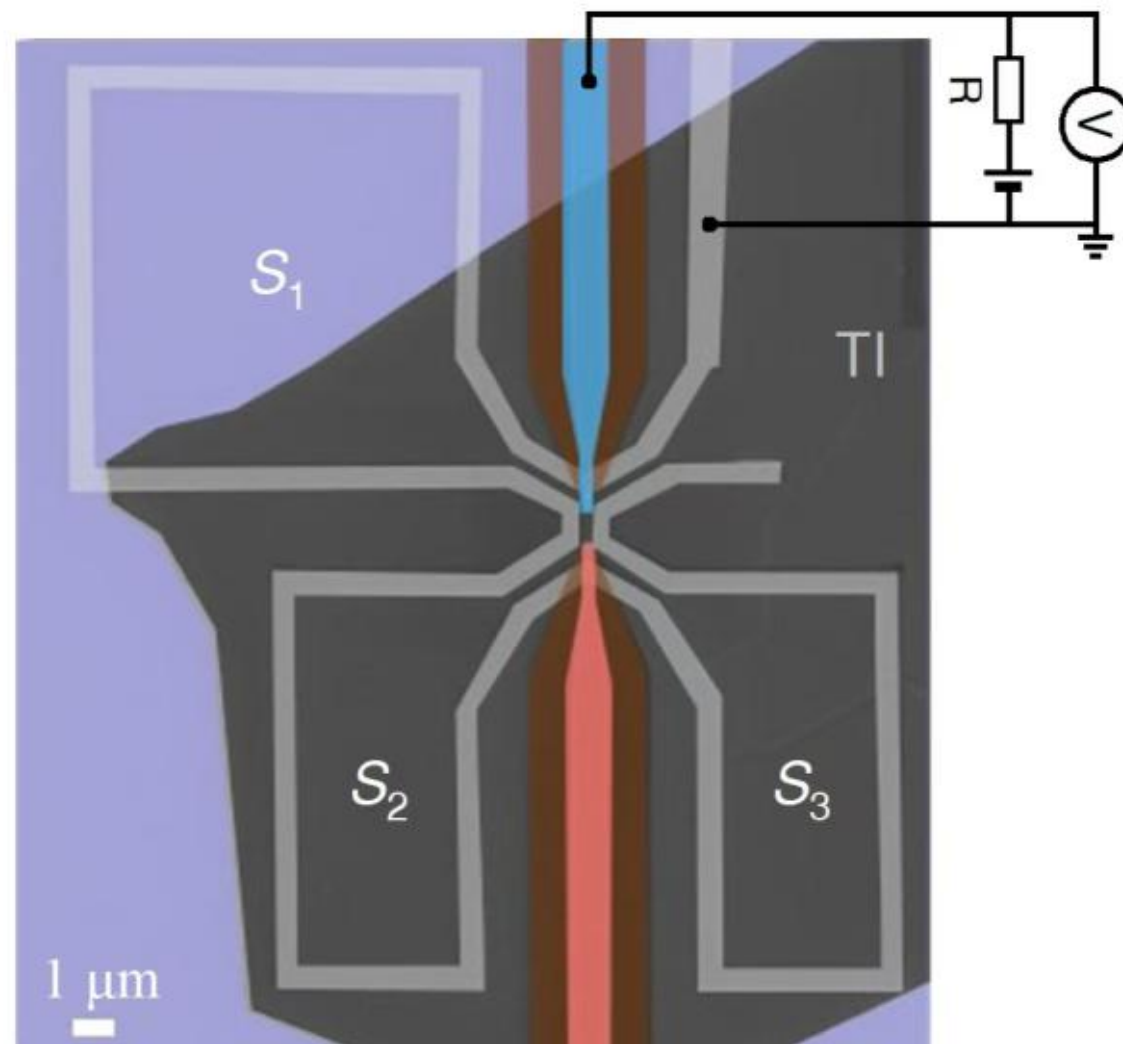

Fig. 1. False-color SEM images of a device containing two Josephson trijunctions separated by 1 μm, constructed on the surface of a Sn-$(Bi,Sb)_2(Te,S)_3$ flake (black) on Si/$SiO_2$ substrate (purple). The status of the trijunctions can be controlled by the magnetic fluxes in the three superconducting loops (gray). In this particular device, the loop areas satisfy $S_1 = 2S_2 = 2S_3$. Two normal-metal electrodes (blue and red) are used to probe the local minigap at the centers of the trijunctions through contact resistance measurement. $Al_2O_3$ pads (brown) are used to isolate the normal-metal electrodes from contacting with the rest part of the device except at the trijunction centers.

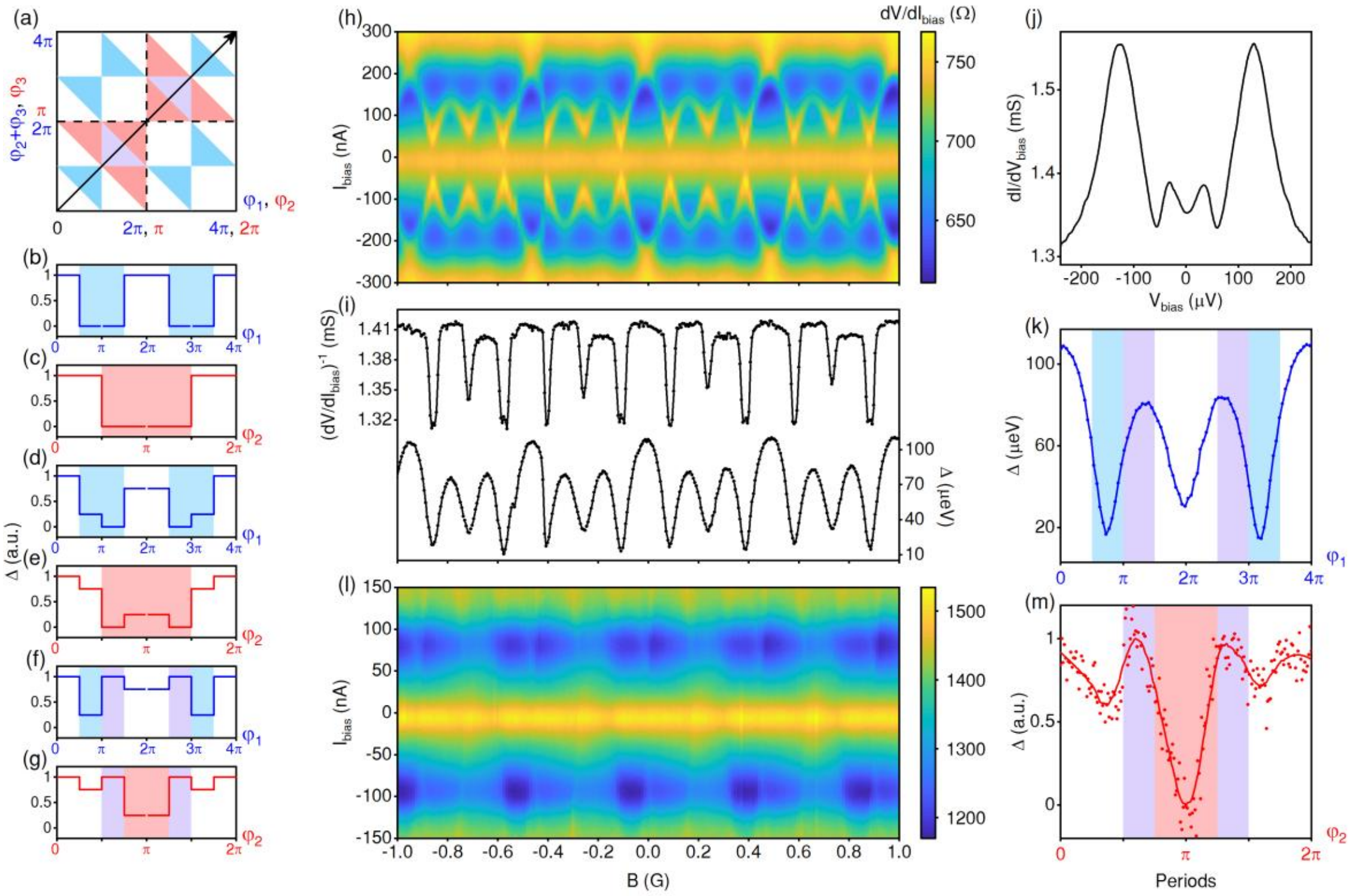

Fig. 2. (a) Fu-Kane's MZM phase diagrams of the upper trijunction (blue coordinates) and the lower trijunction (red coordinates) in the $S_1 = 2S_2 = 2S_3$ configuration. With the variation of a global magnetic field, the status of both trijunctions evolves along the diagonal line. The minigap of the upper/lower trijunction should close in the blue/red regions, except in the overlapped purple regions where the coexistence and coupling of MZMs in the two trijunctions reopen their minigap. (b, c) The expected qualitative line shapes of magnetic field dependence of minigap in the upper and lower trijunctions, respectively, when there is no state penetration from one trijunction's center to the other nor MZM coupling between the two centers. (d, e) With partial state penetration (assuming causing 1/4 change in apparent amplitude of the local minigap), but without MZM coupling. (f, g) With partial state penetration and with MZM coupling. (h) 2D color maps of contact resistance measured by the blue electrode at the upper trijunction's center, as a function of magnetic field and bias current. (i) Correspondence between $(\mathrm{d}V/\mathrm{d}I_{\mathrm{bias}})^{-1}$, the inverse of contact resistance along the horizontal linecut in (h) at $I_{\mathrm{bias}} = -70$ nA (the upper curve), and the minigap (the lower curve) determined by the position of the smaller conductance peaks in (j). (j) Typical bias voltage dependence of contact conductance converted from the data of a vertical linecut in (h) at $B$= 0.73 G. Besides the two big peaks at approximately 130 μeV caused by the induced gap, there are two small peaks representing the ABSs-related minigap whose position oscillates with magnetic field, as shown in (i). (k) Two periods of magnetic field dependence of minigap at the upper trijunction's center, demonstrating a line shape which is in qualitative agreement with the one shown in (f). (l) 2D

color maps of contact resistance measured by the red electrode at the lower trijunction's center, as a function of magnetic field and bias current. (m) Magnetic field dependence of the minigap at the lower trijunction's center obtained from the linecut in (l) at $I_{\mathrm{bias}} = -15$ nA. The data are normalized by setting the maximum as 1 and the minimum as 0. The overall line shape is in qualitative agreement with the one shown in (g).

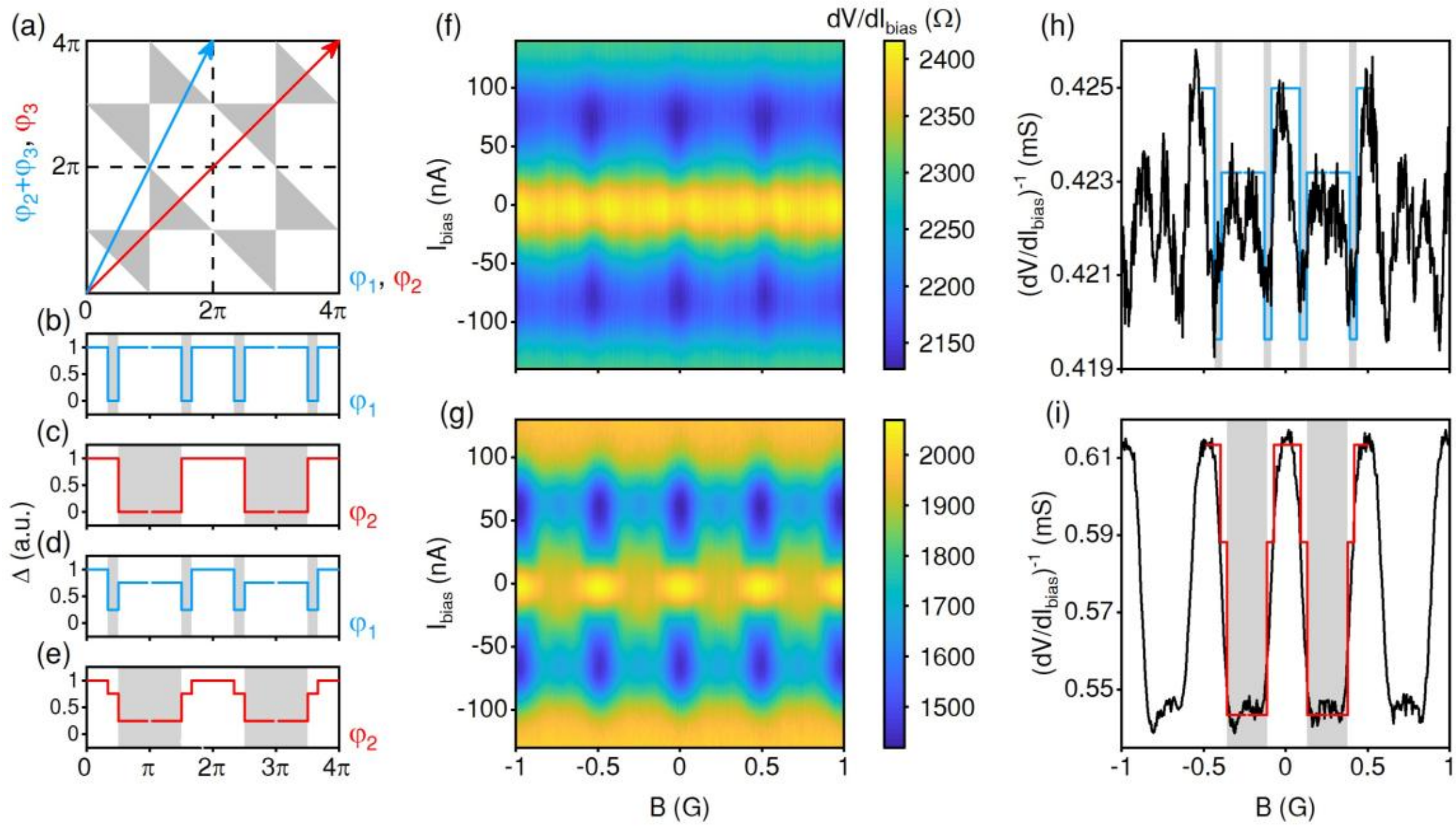

Fig. 3. (a) Fu-Kane's MZM phase diagrams of the upper trijunction (blue coordinates) and the lower trijunction (red coordinates) in the $S_1 = S_2 = S_3$ configuration. With the variation of a global magnetic field, the status of the upper trijunction evolves along the blue line, and the status of the lower trijunction evolves along the red line. The minigap of the two trijunctions will close in the gray regions nonsimultaneously during the evolution. (b, c) The expected qualitative line shapes of magnetic field dependence of minigap in the upper and lower trijunctions, respectively, when there is no state penetration from one trijunction's center to the other. (d, e) The line shapes with partial state penetration, assuming that the penetration causes 1/4 change in apparent amplitude of the local minigap. (f, g) 2D color maps of contact resistance as a function of magnetic field and bias current, measured at the upper and lower trijunction's centers, respectively. (h, i) Magnetic field dependences of $(\mathrm{d}V/\mathrm{d}I_{\mathrm{bias}})^{-1}$ from the horizontal linecuts at 10 nA in (f) and at 30 nA in (g), respectively. Also superimposed in (h, i) are two periods of expected qualitative line shapes of minigap depicted in (d, e), respectively. The gray regions denote the minigap closure/MZM occurring states.

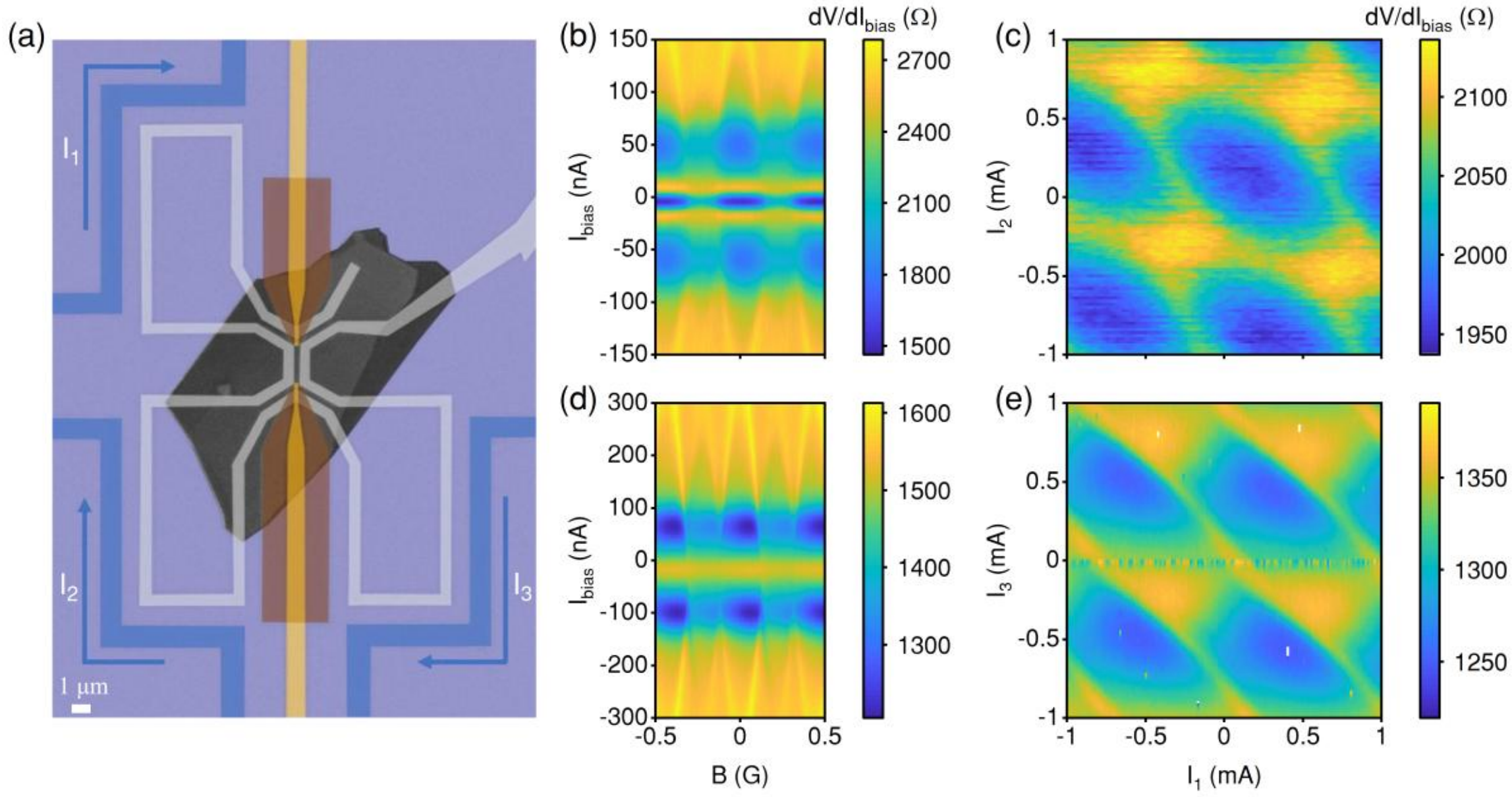

Fig. 4. (a) False-color SEM image of a second type of devices. Three superconducting Al wires (in blue) are added to allow individual adjustment of the local magnetic fluxes in the superconducting loops by currents $I_1$, $I_2$ and $I_3$. The rest of the device is similar to the one shown in Fig. 1(a). (b) 2D color map of contact resistance measured as a function of global magnetic field and bias current at the lower trijunction's center. (c) 2D color map of contact resistance as a function of $I_1$ and $I_2$ measured at $I_{\text{bias}} = -3$ nA. (d) 2D color map of contact resistance measured as a function of global magnetic field and bias current at the lower trijunction's center on another similar device. (e) 2D color maps of contact resistance as a function of $I_1$ and $I_3$ for that device measured at $I_{\text{bias}} = 80$ nA.

# Supplemental Material for "Gap reopening as a possible signature of coupling between Majorana zero modes in Sn-(Bi,Sb)$_2$(Te,S)$_3$-based Josephson trijunctions "

Duolin Wang[1,2], Xiang Zhang[1], Yunxiao Zhang[1,2], Heng Zhang[3], Fucong Fei[3], Xiang Wang[1,2], Bing Li[1,2], Xiaozhou Yang[1,2], Yukun Shi[1,2], Zhongmou Jia[1,2], Enna Zhuo[1,2], Yuyang Huang[1,2], Anqi Wang[1,2], Zenan Shi[1,2], Zhaozheng Lyu[1,2,4,†], Xiaohui Song[1,4], Peiling Li[1], Bingbing Tong[1], Ziwei Dou[1], Jie Shen[1], Guangtong Liu[1,4], Fanming Qu[1,2,4] Fengqi Song[3,†] and Li Lu[1,2,4,†]

[1] *Beijing National Laboratory for Condensed Matter Physics, Institute of Physics, Chinese Academy of Sciences, Beijing 100190, China*

[2] *School of Physical Sciences, University of Chinese Academy of Sciences, Beijing 100049, China*

[3] *College of Physics, Nanjing University, Nanjing 210008, China*

[4] *Hefei National Laboratory, Hefei 230088, China*

† Corresponding authors: lyuzhzh@iphy.ac.cn, songfengqi@nju.edu.cn, lilu@iphy.ac.cn

## 1. Another device in the $S_1 = 2S_2 = 2S_3$ configuration that exhibited the MZM coupling effect

Besides the data shown in Fig. 2 of the main manuscript supporting the existence of inter-trijunction MZM coupling, here we present additional supporting evidence obtained on another device in the $S_1 = 2S_2 = 2S_3$ configuration. Figure S1(a) is the 2D color map of contact resistances measured by the blue probe electrodes as a function of bias current and global magnetic field. Figure S1(b) further shows the $(\mathrm{d}V/\mathrm{d}I_{\mathrm{bias}})^{-1}$ data along the horizontal linecut in Fig. S1(a) at 14 nA. According to the correspondence between the contact conductance and the minigap, the magnetic field dependence of $(\mathrm{d}V/\mathrm{d}I_{\mathrm{bias}})^{-1}$ of this device aligns with the qualitative line shape of minigap depicted in Fig. 2(f) of the main manuscript. The data demonstrate three maxima with in two periods of the upper trijunction, supporting the picture of minigap reopening due to MZM coupling between the two trijunctions.

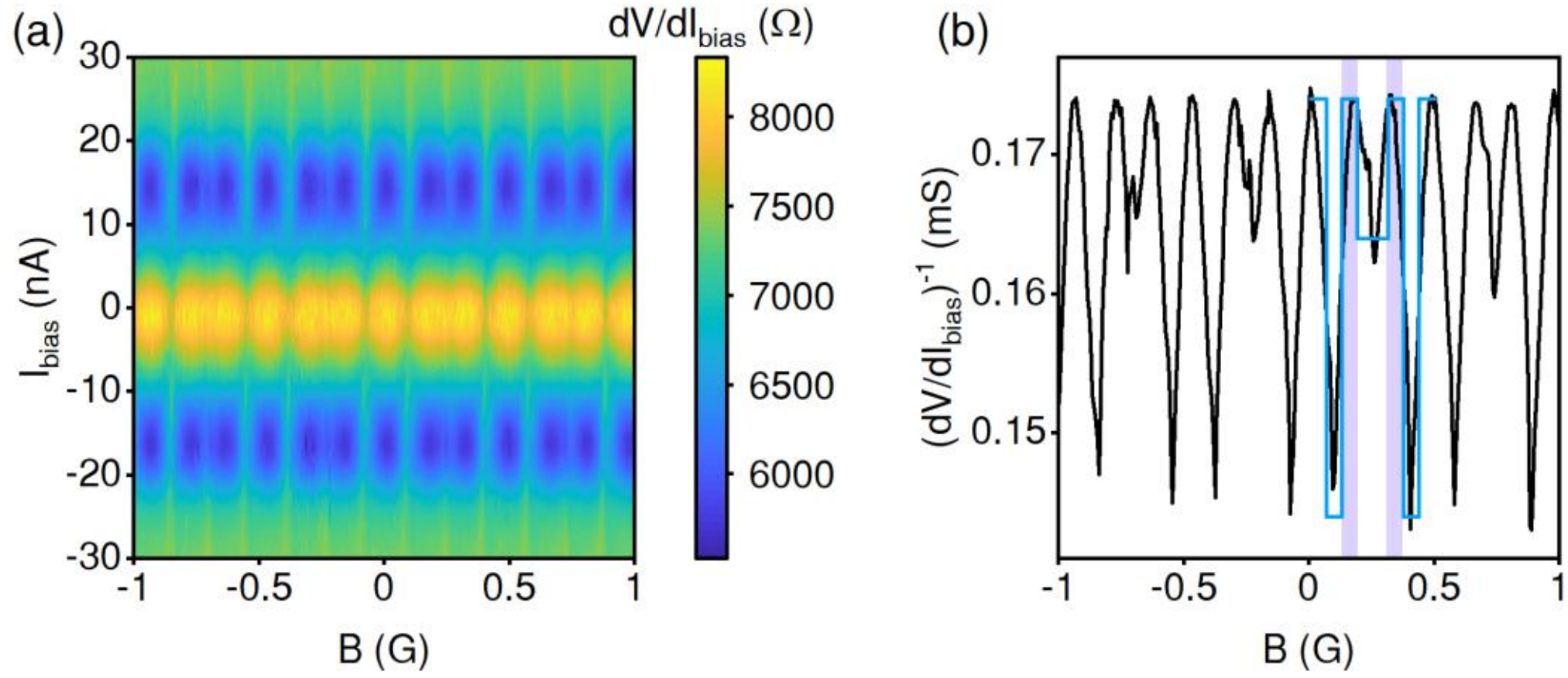

FIG. S1. (a) 2D color map of contact resistances as a function of bias current and global magnetic field, measured by the blue probe electrodes on another device in the $S_1 = 2S_2 = 2S_3$ configuration ($S_1 \approx 90\ \mu\mathrm{m}^2$). (b) The horizontal linecut at 14 nA in (a), plotted as $(\mathrm{d}V/\mathrm{d}I_{\mathrm{bias}})^{-1}$vs. $B$. Also depicted as a blue line are two periods ($\Delta B \approx 0.48$ G) of the expected qualitative line shape of minigap in the upper trijunction [see Fig. 2(f) of the main manuscript], with minigap reopening twice in the purple regions due to MZMs coupling in the two trijunctions, leading to three maxima in two periods.

## 2. Conversion of $\mathbf{d}V/\mathbf{d}I_{\mathbf{bias}}$ data to $\mathbf{d}I/\mathbf{d}V_{\mathbf{bias}}$ data

While discussing Andreev Bound State-related physics, it would be more appropriate to use the $\mathrm{d}I/\mathrm{d}V_{\mathrm{bias}}$ data instead of the $\mathrm{d}V/\mathrm{d}I_{\mathrm{bias}}$ data. For this purpose, we can convert the vertical linecuts, namely the $\mathrm{d}V/\mathrm{d}I_{\mathrm{bias}} - I_{\mathrm{bias}}$ curves, in Fig. 2(h) of the main manuscript at each magnetic field to $V - I_{\mathrm{bias}}$ curves through integration, then further convert the data to $\mathrm{d}I/\mathrm{d}V_{\mathrm{bias}} - V_{\mathrm{bias}}$ curves through differential. One of the converted curves is shown in Fig. 2(j) of the main manuscript. Shown in Fig. S2 below is the whole 2D map of $\mathrm{d}I/\mathrm{d}V_{\mathrm{bias}}$ data converted form Fig. 2(h) of the main manuscript.

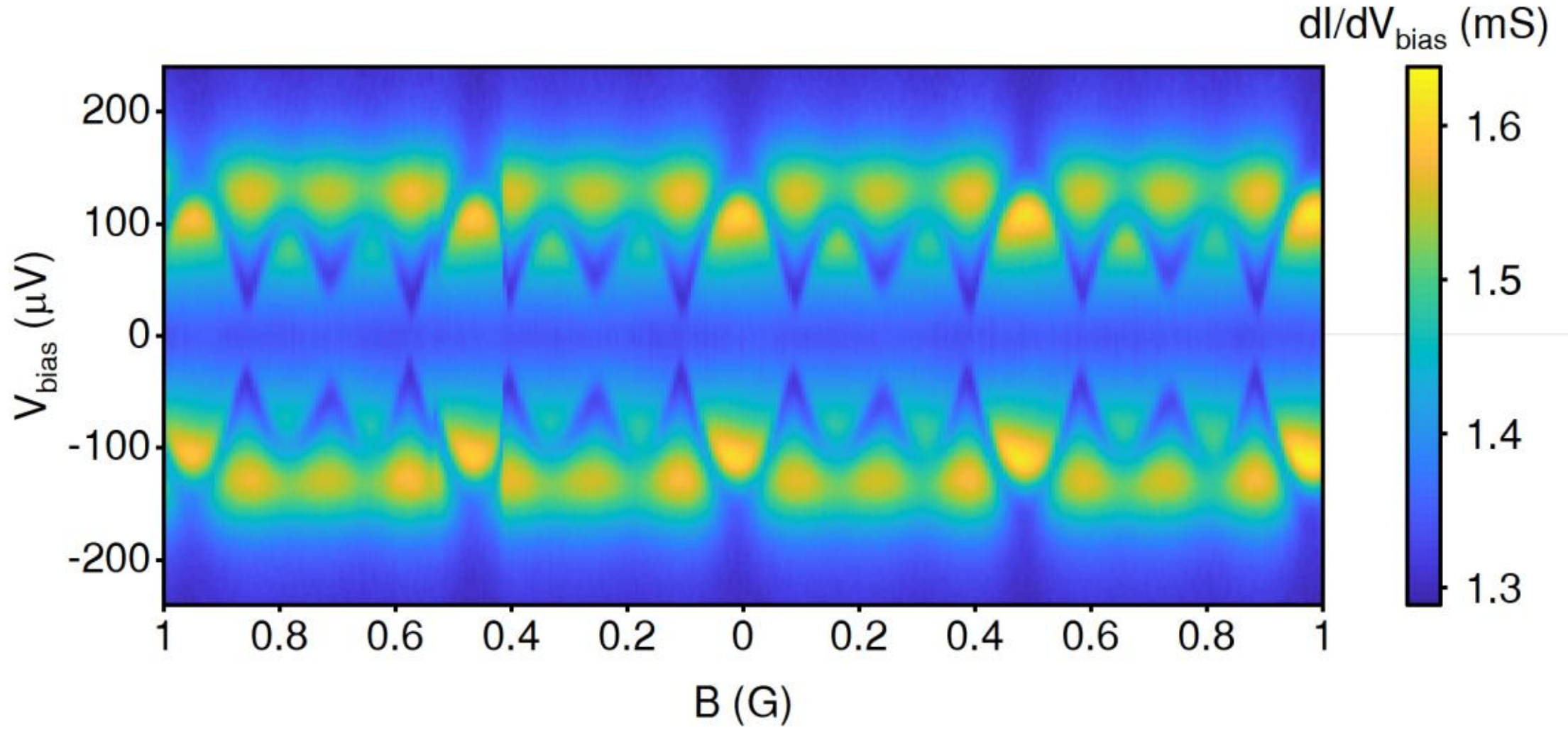

FIG. S2. 2D map of $\mathrm{d}I/\mathrm{d}V_{\mathrm{bias}}$ data converted form Fig. 2(h) of the main manuscript.

Because the magnetic field only causes a small fraction of modulation on either $\mathrm{d}V/\mathrm{d}I_{\mathrm{bias}}$ or $\mathrm{d}I/\mathrm{d}V_{\mathrm{bias}}$, there is essentially no fundamental difference between the patterns in the two 2D maps. And, the $(\mathrm{d}V/\mathrm{d}I_{\mathrm{bias}})^{-1}$ data obtained along a linecut at a fixed bias current in the $\mathrm{d}V/\mathrm{d}I_{\mathrm{bias}}$ 2D map should be comparable to the $\mathrm{d}I/\mathrm{d}V_{\mathrm{bias}}$ the data along a properly chosen linecut at a fixed bias voltage in converted $\mathrm{d}I/\mathrm{d}V_{\mathrm{bias}}$ 2D map.

### 3. A direct comparison between the phase diagram and the experimental data showing the results of minigap reopening

Shown in Fig. S3 are the phase diagram for the $S_1 = 2S_2 = 2S_3$ configuration, aligned together with the data shown in Figs. (k, m) of the main manuscript. It allows direct comparison between the phase diagram and the data, demonstrating the happening of minigap reopening in the purple-colored regions.

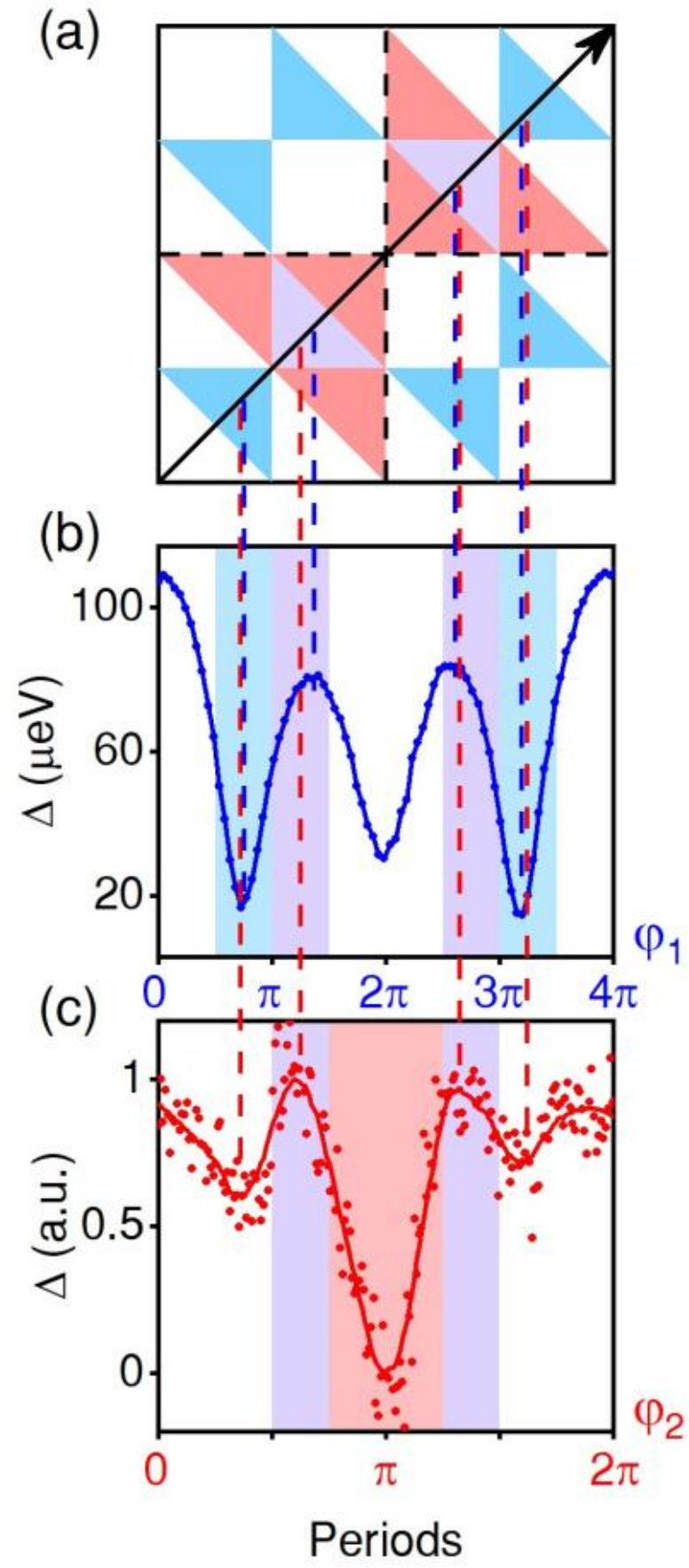

FIG. S3. (a) Double trijunctions' MZM phase diagrams in the $S_1 = 2S_2 = 2S_3$ configuration. (b) Two periods of magnetic field dependence of minigap at the upper trijunction's center, demonstrating three minigap lobes. (c) One period of normalized minigap at the lower trijunction's center. The minigaps reopen and maximize in the purple-colored regions of the phase diagrams where MZMs coexist.

## 4. An example of failure to observe the coupling effect in the $S_1 = 2S_2 = 2S_3$ configuration

As we have mentioned in the main manuscript, the coupling effect was observed on only two out of more than ten devices in the $S_1 = 2S_2 = 2S_3$ configuration. More often, no obvious coupling effect could be recognized. Figures S4(a, b) present the typical data of contact resistance measured by the upper (blue) electrodes on two separate devices, whose magnetic field dependence of minigap can be described by the qualitative line shape in the absence of coupling, as depicted in Fig. 2(b).

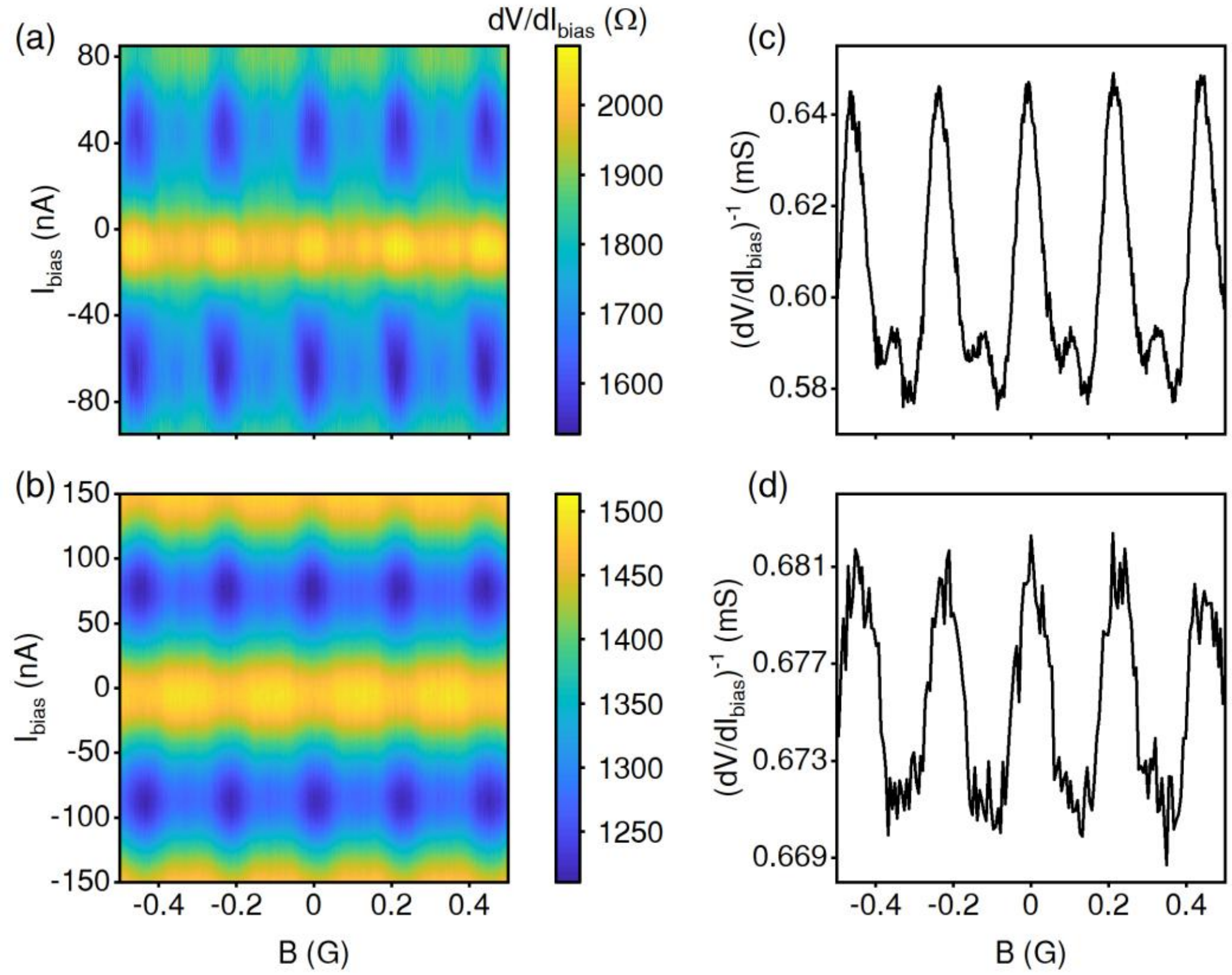

FIG. S4. (a, b) 2D color maps of contact resistance as a function of global magnetic field and bias current, measured by the upper (blue) electrodes on two separate devices in the $S_1 = 2S_2 = 2S_3$ configuration ($S_1 \approx 90$ μm$^2$). (c) The horizontal linecut at -60 nA in (a), exhibiting a period of $\Delta B \approx 0.24$ G . (d) The horizontal linecut at -10 nA in (b), exhibiting a period of $\Delta B \approx 0.22$ G. No obvious coupling effect can be recognized.